\documentclass{aa}
\usepackage{graphics,epsfig,lscape}
\usepackage{natbib}

\bibpunct{(}{)}{;}{a}{}{,}

\def\cm-2{cm$^{-2}$}

\def\n253{\object{NGC~253}}
\def\m33{\object{M~33}}
\def\mx7{\object{M~33~X$-$7}}
\def\x7{\hbox{X$-$7}}

\begin{document}

\title{M31N 2005-09c: a fast \ion{Fe}{ii} nova in the disk of M~31}

   \author{D.~Hatzidimitriou
   \inst{1} \and
       P.~Reig\inst{2} \and
       A.~Manousakis \inst{1} \and
       W.~Pietsch\inst{3} \and
       V.~Burwitz\inst{3} \and
       I.~Papamastorakis \inst{1}}

       \institute{University of Crete, Department of Physics,P.O.Box 2208, 71003,
       Heraklion, Greece \and IESL, Foundation for Research and Technology, 71110,
            Heraklion, Greece  \and Max-Planck-Institut f\"ur extraterrestrische Physik,
           85741 Garching, Germany}

     \offprints{D.~Hatzidimitriou, e-mail: {\tt dh@physics.uoc.gr}}

   \date{Received; accepted }

    \abstract
     {Classical novae are quite frequent in M~31. However, very few spectra
     of M31 novae have been studied to date, especially during the early decline phase.}
     {Our aim is to study the photometric and spectral evolution of a
     M~31 nova event close to outburst.}
      {Here, we present photometric and spectroscopic observations of M31N 2005-09c, a classical
      nova in the disk of M~31, using the 1.3m telescope of the Skinakas Observatory in Crete
     (Greece), starting on the 28th  September, i.e. about 5 days after outburst,
     and ending on the 5th October 2005, i.e. about 12 days after
     outburst. We also have supplementary photometric observations
     from the La Sagra Observatory in Northern Andaluc\'{\i}a, Spain, on September 29 and 30,
     October 3, 6 and 9 and November 1, 2005. The wavelength range covered by the spectra is
     from 3565 \AA\ to 8365 \AA. The spectra are of high S/N allowing the study
     of the evolution of the equivalent widths of the Balmer lines, as well as the
     identification of non-Balmer lines.}
     {The nova displays a typical early decline spectrum that is characterized by many weak
     \ion{Fe}{ii} multiplet emissions. It is classified as a P$_{fe}$ nova.
     From the nova light curve, we have also derived its speed class, $t_2=14\pm2.5$ days.
      As the nova evolved the Balmer lines became stronger and narrower. The early decline of
     the expansion velocity of the nova follows a power law in time
     with an exponent of $\simeq-0.2$.}{}

   \authorrunning{Nova in M31}
   \maketitle

\section{Introduction}
Classical novae are a sub-class of cataclysmic variables, that
exhibit outburst magnitudes of 10-20 mag, reaching peak
luminosities as high as $M_V=-9.5$ for the fastest ones (Warner
1989). Nova outbursts are due to thermonuclear runaways of the
hydrogen-rich material that has been accreted onto the white dwarf
primary.

The properties of novae may vary between different galaxies (see
e.g. Shafter \& Irby 2001). This is an important issue that needs
to be clarified before novae can be considered as reliable
distance indicators (e.g. Della Valle \& Livio, 1995). Moreover,
the systematic study of novae in different galaxies in the Local
Group is important for improving our understanding of the eruption
process. M~31 is an excellent target in this context, due to its
proximity and relatively high nova rate.

Novae have been observed in M~31 ever since the pioneering work of
Hubble (1929). Several surveys have been published since (cf
Darnley et al. 2004, 2005; Pietsch et al. 2005, and references
therein), leading to the discovery of over 470 novae in this
galaxy. According to the estimates of Darnley et al. (2005), from
the POINT-AGAPE survey, the global nova rate in M~31 is
$65^{+16}_{-15}yr^{-1}$.

Despite the large number of novae discovered in M~31 and their
potential importance in achieving a deeper understanding of nova
outburst processes in external galaxies, very few spectra of M~31
novae have been published to date. Tomaney \& Shafter (1992)
provided the only extensive spectroscopic study of M~31 novae to
date, for 9 objects lying in the M~31 Bulge. They further analyzed
one of these objects, a remarkably slow nova (Tomaney \& Shafter,
1993), which had also been studied spectroscopically earlier
(Cowley \& Starrfield 1987).

On September 23, 2005, Quimby et al.(2005) reported the discovery
of an optical transient in the direction of M31, at RA$ =
00^h38^m54.63^s$, DEC$ = +40^o27'34.2''$ (J2000). The location of
the transient is shown in Figure 1, and it appears to be related
to the disk of M~31. The detection was based on unfiltered CCD
images taken on September 22.17 UT and September 23.21 UT, with
the 0.45m ROTSE-IIIb telescope at the McDonald Observatory. The
magnitude of the transient was about 16.3 mag on the first
observation and 16.0 mag on the second. The object was not
detected in ROTSE-IIIb data from September 20.18 UT (limiting mag
of about 17.6).  The transient was spectroscopically confirmed to
be a classical nova (Reig et al. 2005). Following the Central
Bureau for Astronomical Telegrams (CBAT) nomenclature
(http:$//$www.cfa.harvard.edu$/$iau$/$CBAT\_M31.html), we shall
refer to this nova as M31N 2005-09c.

%\begin{figure}
%\resizebox{\hsize}{!}{\rotatebox{0}{\includegraphics{fig1.eps}}}
%\caption[]{The location of the nova M31N 2005-09c at RA$ =
%00^h38^m54.63^s$, DEC$ = +40^o27'34.2''$ (J2000), is marked on the
%M~31 image taken from Sky View and  based on photographic data
%obtained using the Oschin Schmidt Telescope on Palomar Mountain. }
%\end{figure}

\begin{figure}
\resizebox{\hsize}{!}{\rotatebox{0}{\includegraphics{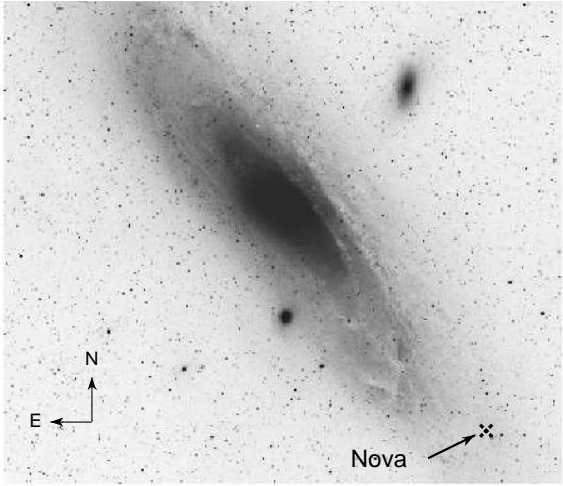}}}
\resizebox{\hsize}{!}{\rotatebox{0}{\includegraphics{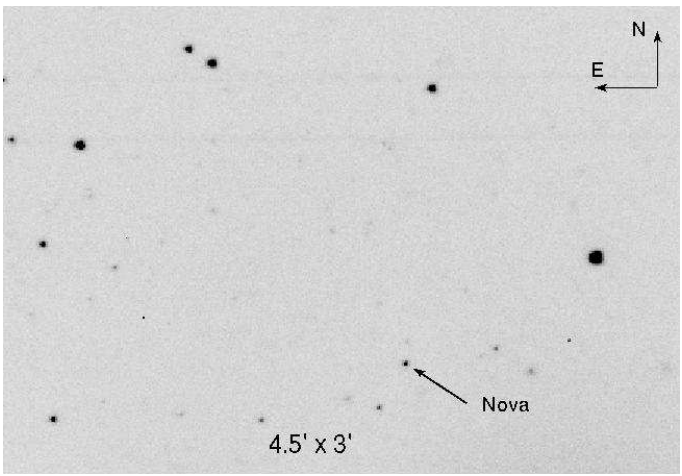}}}
\caption[]{Finding chart of nova M31N 2005-09c. Its location at
RA$ = 00^h38^m54.63^s$, DEC$ = +40^o27'34.2''$ (J2000), is marked
on the M~31 image taken from NASA's Sky View Virtual Observatory,
based on photographic data obtained using the Oschin Schmidt
Telescope on Palomar Mountain. }
\end{figure}

In this paper we have carried out optical observations of M31
2005-09c, in order to monitor  its photometric and spectral
variability. The spectroscopic observations started on September
28, 2005, i.e., 5 days after outburst, and ended on October 5,
2005, i.e., 12 days after outburst, while the photometric follow
up of the nova ended on November 1, 2005.

In Section 2, we describe the optical observations, in Section 3
the data reduction, in Section 4 we present and analyze the
results. Finally, in Section 5 we summarize our conclusions.

\section{Observations}
The spectroscopic observations used in this study were carried out
during four nights, on September 28 and 29, 2005 and on October 3
and 5, 2005, using the 1.3-m Ritchey-Cretien telescope at Skinakas
Observatory, located on the island of Crete (Greece).

The telescope was equipped with a 2000x800 ISA SITe CCD camera and a
1302 lines/mm grating, blazed at 5500 \AA. Spectra were obtained in
four overlapping wavelength regions in order to include as many
spectral lines of interest as possible. The overall spectral range
covered was from 3565-8365 \AA. The nominal dispersion was 1.04
\AA/pixel. Exposure times ranged from 600 to 3600~s, depending on
brightness of the object and on weather and seeing conditions. Each
on-target exposure was followed by an arc calibration exposure. The
log of the observations is given in Table 1.

The acquisition data which were unfiltered, were used to derive
approximate magnitudes of the nova, as will be described in detail
in the next section.

\begin{table}
\begin{center}
\caption{Log of spectroscopic observations of M31N 2005-09c}

\begin{tabular}{cccc}
\hline\hline\noalign{\smallskip}
Date &  JD  & Spectral           & Exposure  \\
     &  2453600.0$+$     & Range(\AA)         & Time (s) \\
\hline\noalign{\smallskip}
28sep05 & 42.307  &4750-6825 & 600\\
        & 42.315  &4750-6825 &1800\\
        & 42.338  &4750-6825 &3000\\
29sep05 &  43.304 &3565-5675 &3000\\
        &  43.417 &3565-5675 &3500\\
        &  43.491 & 3565-5675&3500\\
        &  43.349 &5090-7160 &3000\\
        &  43.542 & 5090-7160 &3000\\
        &  43.260 &4750-6825&3000\\
03oct05 &47.390  &4750-6825 &3300\\
        & 47.446 &6315-8365 &3300\\
05oct05 &49.264  &3565-5675 &3600\\
        & 49.312 &3565-5675 &3600\\
        & 49.370&4750-6825 &3600\\
        & 49.414& 4750-6825 &3600\\
        & 49.471&6315-8365 &3600\\
\hline\hline\noalign{\smallskip}
\end{tabular}
\end{center}
\end{table}

We also have some complimentary photometric observations from the
La Sagra Observatory (run by Observatorio Astronomic de Mallorca,
hereafter, OAM) in north-east Andaluc\'{\i}a, Spain. These
observations were obtained on September 29 and 30, October 3, 6
and 9, and November 1, 2005, using a Celestron CGE1400 14 inch
telescope with a Hyperstar focal reducer and a 10XME camera with a
KODAK KAF-3200 ME CCD chip. The pixel scale was 2.13$''$ per
6.8$\mu$ pixel, with a total field of view of 77.4$'$ by 52.1$'$.
The images were white light images (i.e. no filter was used). The
dates and times of the observations are listed in Table 2 along
with the derived magnitudes of the nova (see next section).

\section{Data Reduction}

\subsection{Photometry}
\subsubsection{Skinakas Observatory}
The "white light magnitudes" of the nova M31N 2005-09c were
converted to R magnitudes by comparing them against the R
magnitudes of USNO-B1 (Monet et al. 2003) stars in the acquisition
fields. This was achieved by using 9 bright USNO-B1.0 stars,
present in the acquisition field, as secondary standards. For each
image, a separate calibration curve was produced from the
instrumental magnitudes of the 9 stars and the corresponding
USNO-B1.0 {\em R} magnitudes \footnotetext{The USNO-B1 magnitudes
are tied to the Jonhson-Kron-Cousins system (cf Bucciarelli et al.
2001), therefore the quoted R magnitude is "Cousins R".}. The
calibration curve was approximated by a linear fit, from which we
then derived the {\em R} magnitude and associated error  of the
nova. These estimates are presented in the upper panel of Table 2.

 One important caveat regarding the accuracy of the derived
R-magnitudes, is that the nova is an emission line object,while the
USNO calibration stars are mostly "normal" stars. In order to at
least partly address this problem, we compared the colour
distribution of known novae (van den Bergh \& Younger 1987 for novae
near maximum and Szkody 1994 for older novae) to the colour
distribution of the calibrating stars. The colour range of the
calibrating stars covers that of novae (near maximum and older
ones), although the  peaks of the colour distributions of the
calibrating stars and of novae are somewhat different, with the
novae being bluer in the mean than the calibrating stars. There is a
small colour term in the magnitude residuals of the calibrators,
indicating that a correction of 0.09 mag should be applied to bluer
objects. This correction falls well within the magnitude error
margins quoted in Table 1. Thus, the size of the R-magnitude error
that we derived indeed should include the error due to the unknown
spectral type of the specific nova. In any case, it must be
emphasized that all of the nova data used to provide the lightcurve
discussed in Section 4.1, including the La Sagra (see below) and the
Quimby et al. (2005) data are calibrated against USNO-R magnitudes.
Therefore, even if there is a zero-point error in the lightcurve,
the speed class derived from the lightcurve should be valid.

\subsubsection{La Sagra Observatory, OAM}

For each night, a median image was created from 15 individual
images of 60 sec exposure each. These median images  were then
reduced and analyzed using the Astrometrica program (by Herbert
Raab, www.astrometrica.at), which automatically selects stars and
compares them to the USNO-B1.0 catalogue, in order to provide
calibration curves for the photometry. The resulting {\em R}
magnitudes and associated errors for the nova M31N 2005-09c are
listed in the lower panel of Table 2.

For the nights of September 29, 2005 and October 3, 2005, there
are magnitude estimates for the nova from both observatories. In
the former, the difference between the magnitudes
($R_{Skinakas}-R_{OAM}$) obtained closest in time is $0.30\pm0.22$
mag, while in the latter, $-0.33\pm0.20$ mag. Thus, there is no
evidence of a systematic difference between the photometry from
the two telescopes, and the random errors are within the expected
range.

As can be seen in Table 2, for the nights of September 15, 2005
and November 1, 2005, only lower limits  for the magnitude of the
nova could be derived. Actually, all available OAM images of the
field were examined, both before the outburst, going back to
September 25, 2005 and after the last date shown in Table 2
(November 1, 2005), until the December 1, 2005. The object was not
detected on any of these images (i.e. it was fainter than
R$\simeq$20.5 mag).

\begin{table}
\begin{center}
\caption{Photometry from Skinakas Observatory and from the La
Sagra Observatory}

\begin{tabular}{ccccc}
\hline\hline\noalign{\smallskip}
Date & JD & {\em R} \\
  &2453600.0$+$&mag\\
\hline\noalign{\smallskip}
\multicolumn{3}{c}{{Skinakas Observatory}}\\
\hline\noalign{\smallskip}
28.09.05&42.303&  17.44$\pm$0.22\\
29.09.05&43.251 & 17.49$\pm$0.14\\
       &43.407 & 17.51$\pm$0.18\\
       &43.481 & 17.48$\pm$0.22\\
03.10.05 &47.262 & 17.54$\pm$0.17\\
       &47.316 & 17.47$\pm$0.12\\
05.10.05 &49.255 & 17.65$\pm$0.20\\
       &49.366 & 17.72$\pm$0.24\\
       &49.467 & 17.55$\pm$0.21\\
\hline\noalign{\smallskip}

\multicolumn{3}{c}{{La Sagra Observatory}}\\
\hline\noalign{\smallskip}
15.09.05  &29.856  & $>$20.5\\
29.09.05  &43.190   & 17.18$\pm$0.14\\
30.09.05  &44.127   & 17.44$\pm$0.14\\
03.10.05   &47.966  & 17.80$\pm$0.16\\
06.10.05  &50.184   & 17.73$\pm$0.18\\
09.10.05  &53.012   & 18.47$\pm$0.22\\
01.11.05  &76.949   & $>$20.2\\
\hline\hline\noalign{\smallskip}

\end{tabular}
\end{center}
\end{table}

\subsection{Spectroscopy}
The reduction of the spectra was performed using the {\em
STARLINK} Figaro package (Shortridge et al. 2001). The frames were
bias subtracted, flat fielded and corrected for cosmic ray events.
The 2-D spectra were subsequently sky subtracted using the {\em
POLYSKY} command. A spatial profile was then determined for each
2-D spectrum, and, finally, the object spectra were optimally
extracted  using the algorithm of Horne (1986), with the {\em
OPTEXTRACT} routine. Arc spectra were then extracted from the arc
exposures, using exactly the same profiles as for the
corresponding object spectra. The arc spectra were subsequently
used to calibrate the object spectra.

 The full width at half maximum (FWHM) of the arc lines were also
used to estimate the instrumental broadening (resolution) of the
spectral lines. The instrumental broadening depends on the position
of the line on the CCD chip. When the line is near the edge the
instrumental width is wider. Typical values are 6 \AA\ at the edges
and about half that value in the center of the chip. The quoted
values of the FWHM of the spectral lines reported here were
corrected for such an effect, taking into account their position on
the chip.

\section{Results}

\subsection{Lightcurve}
In Figure 2, we present an estimate of the light curve of the nova,
based on the magnitudes of Table 2 and on the magnitude estimates of
Quimby et al.(2005), mentioned in Section 1. The former are
presented as filled circles (from Skinakas Observatory) and
asterisks (La Sagra Observatory, OAM), along with their associated
errors, as given in Table 2. The Quimby et al. values are shown as
open circles. The ROTSE-IIIb data on which these values were based
were unfiltered (roughly 400-800nm). The magnitudes reported in
Quimby et al. (2005) were calibrated against the USNO-A1.0 R band
(Quimby 2006, private communication). The arrows indicate the lower
detection limits of the 20th of September observation of Quimby et
al. (limiting magnitude 17.6, grey arrow), and of the lower
detection limits mentioned in Table 2, from OAM (black arrows).

It appears that the nova had a magnitude around $R\simeq16.0$ mag
near maximum light, which corresponds to an absolute magnitude 
of $M_R\simeq$ -8.6, using $(m-M)_V=24.47\pm0.07$ and E(B-V)=0.06
from Holland (1998) and the conversion $A_R/A_V=0.748$ from Rieke \&
Lebofsky (1985).

The "speed class" of a classical nova is often used to describe the
overall timescale of an eruption and to classify a nova. The speed
class, $t_2$, is the number of days that a nova takes to fade by two
magnitudes below maximum light (e.g. Darnley et al. 2005). According
to the light curve of Figure 2, the nova M31N 2005-09c has faded by
about two magnitudes in $t_2=14\pm2.5$ days (it was about 16th mag,
on the 23rd September and about 18 mag between the 6th and 9th
October). The error is estimated on the basis of the quoted errors
in R and the frequency of observations.

Using the recent maximum magnitude versus rate of decline (MMRD)
relationship  of Downes \& Duerbeck (2000),
$M_V=(-11.32\pm0.44)+(2.55\pm0.32)log(t_2)$,  we estimate that for
the derived rate of decline of $t_2=14\pm2.5$ days,  $M_V$ at
maximum would be expected to be around $\simeq -8.4\pm0.6$. Our
estimate of $M_R$ is  $\simeq$ -8.6, i.e. very close to the value
derived from the MMRD relationship for $M_V$. Note, that given that
novae close to maximum are blue ($(B-V)_o\simeq0.23$; van den Bergh
\& Younger 1987, Downes \& Duerbeck 2000), $(V-R)_o$ is expected to
be small,  well within the quoted errors.

\begin{figure}
\resizebox{\hsize}{!}{\rotatebox{0}{\includegraphics{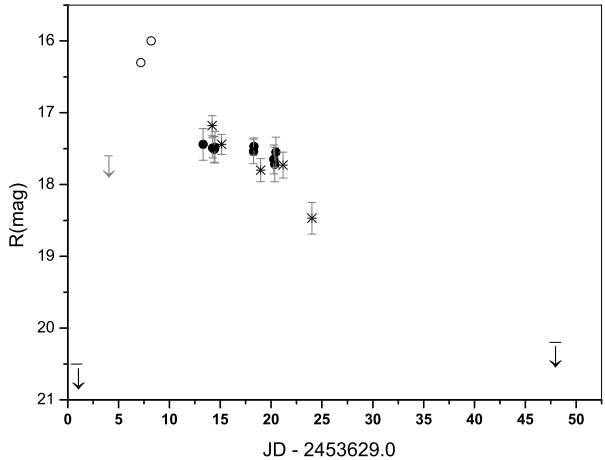}}}
 \caption[]{The light
curve of the nova between September 15, 2005 and November 1, 2005.
Filled circles correspond to the Skinakas measurements and
asterisks to the La Sagra measurements, summarized in Table 2. The
down-arrows show lower limits of the nova's magnitude. Open
circles correspond to approximate magnitudes given by Quimby et
al. 2005. }
\end{figure}

\subsection{Line identification}
Figure 3 shows examples of spectra obtained for the transient, in
three different wavelength regions. The top panel shows the
spectrum from 3700-5675\AA~~(average of three spectra obtained on
September 29), the middle panel from 4750-6825\AA~~(average of two
spectra obtained on September 29, 2005) and the bottom panel from
6315-8365\AA~~(one observation obtained on October 3, 2005). All
spectra have been normalized to unity.

\begin{figure}
\resizebox{\hsize}{!}{\rotatebox{0}{\includegraphics{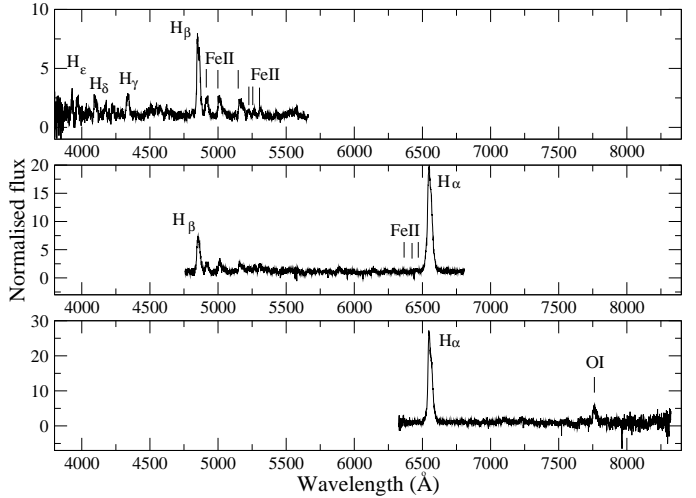}}}
\caption[]{The optical spectrum of the nova obtained in three
different wavelength regions, covering 3700-8200\AA. Top and
middle panels correspond to the average of three and two
observations respectively, obtained on September 29, 2005, while
the bottom panel corresponds to one near-IR observation obtained
on October 3, 2005. All spectra have been normalized to unity. }
\end{figure}

The lines of the Balmer series up to H$\epsilon$ appear strongly in
emission. The spectra are also characterized by \ion{Fe}{ii} emission,
forbidden and permitted. The lines identified are listed in Table 3. Column
1 gives the line identification, with the multiplet number in brackets and
forbidden transitions indicated by square brackets. Column 2 gives the
theoretical rest wavelength of the line (from Thackeray, 1952) and Column 3
the observed wavelength of the line.

Most of the \ion{Fe}{ii} lines lie between the H$\alpha$ and
H$\beta$ lines. Figure 4 shows the spectrum of the nova in this
region, i.e. between 4800 and 6400\AA, with the identified lines
clearly marked.

Apart from the Balmer and \ion{Fe}{ii} lines, two more lines were
also seen in emission, namely, \ion{Na}{iD}
$\lambda5894$\AA~~(shown in Figure 4) and
\ion{O}{i}$\lambda7773$\AA~~(shown in Figure 2).

In conclusion, the nova displays a typical early decline spectrum
that is characterized by many weak \ion{Fe}{ii} multiplet
emissions. Following the Tololo Nova Spectral Classification
System (Williams et al. 1991, Williams et al. 1994), we classify
the nova as a P$_{fe}$ nova. The nova is undoubtedly in the
permitted lined phase (P), as the strongest non-Balmer emission is
a permitted transition. The subclass is assigned from the
strongest non-Balmer line in the spectrum, which is clearly
\ion{Fe}{ii} $\lambda5018$~~\AA,~~ $\lambda5169$~~\AA. The
presence of the \ion{O}{i} line at 7773~~\AA, ~~ suggests that the
nova is close to the \ion{O}{i} flash, which is expected at about
2.5 mag below maximum light (e.g. Tomaney \& Shafter 1992). On the
spectrum of October 5, 2005, there is clear indication of the
\ion{O}{i} $\lambda6300$ ~~\AA,~~ line. This line is expected (at
least for galactic classical novae) to "turn on" at 2.6 mean
magnitudes from maximum (Tomaney \& Shafter 1992).

\begin{table}
\begin{center}
\caption{Identified emission lines }

\begin{tabular}{rrr}
\hline\hline\noalign{\smallskip}
ID$^1$&  $\lambda$ $_o$ (\AA) & $\lambda$ $_{obs}$ (\AA) \\
\hline\noalign{\smallskip}

\ion{Fe}{ii}(3)    & 3930.3 & 3931.3\\
H$\epsilon$& 3970.1 & 3970.0\\
H$\delta$&4101.7& 4101.0\\
H$\gamma$&4340.5& 4341.2\\
$[$\ion{Fe}{ii}$]$(20) & 4852.7 & 4853.2\\
H$\beta$& 4861.3 & 4861.4\\
\ion{Fe}{ii}(42)& 4923.9 & 4922.9\\
\ion{Fe}{ii}(42)& 5018.4 & 5018.2\\
\ion{Fe}{ii}(42)& 5169.0 & 5168.3\\
\ion{Fe}{ii}(49)& 5234.6 & 5235.2\\
$[$\ion{Fe}{ii}$]$(18)& 5268.9 & 5269.8\\
$[$\ion{Fe}{ii}$]$(18)& 5273.4 & 5276.4\\
\ion{Fe}{ii}(41)& 5284.1 & 5283.1\\
\ion{Fe}{ii}(49)& 5327.1 & 5326.5\\
\ion{Fe}{ii}   & 5894.1 & 5894.6\\
\ion{Fe}{ii}(74)& 6147.7 & 6150.0\\
\ion{Fe}{ii}(74)& 6238.4 & 6241.6\\
H$\alpha$& 6562.8 & 6560.4\\
\ion{O}{i}        & 7773.0 & 7774.8\\
 \hline\hline\noalign{\smallskip}
\end{tabular}
\end{center}
 $^1$ {\tiny Suggested identification is followed by the multiplet number in
  brackets. Forbidden lines are distinguished by square brackets.}

\end{table}

\begin{figure}
\resizebox{\hsize}{!}{\rotatebox{0}{\includegraphics{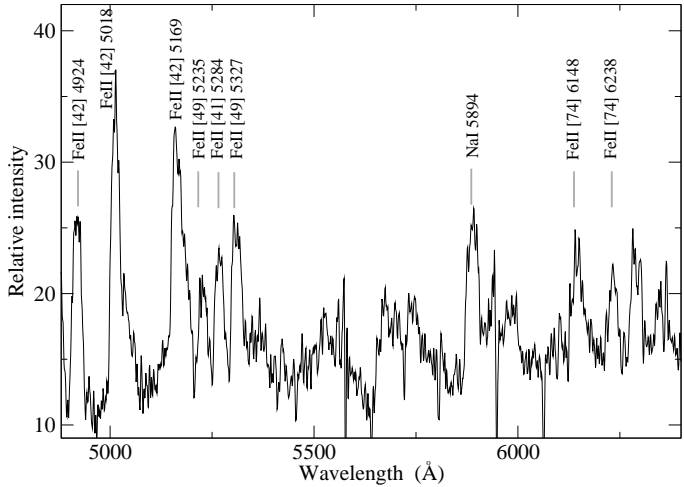}}}
\caption[]{Identification of \ion{Fe}{ii} lines between 4800-6400
\AA. }
\end{figure}

\begin{figure}
\resizebox{\hsize}{!}{\rotatebox{0}{\includegraphics{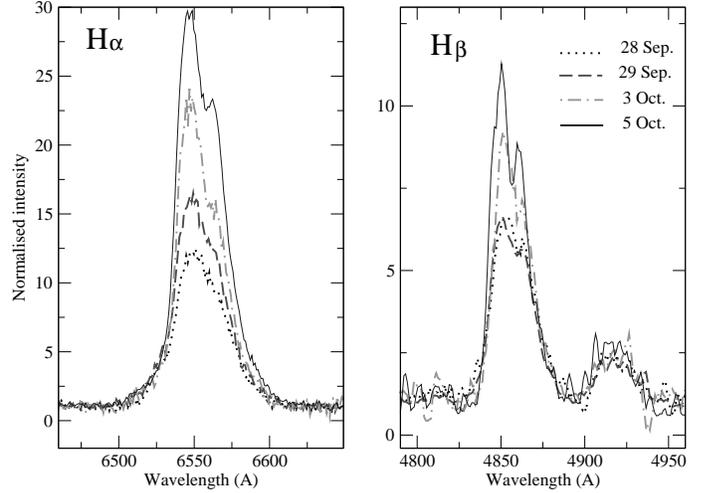}}}
\caption[]{The evolution of the intensity and profile of the
H$\alpha$ and H$\beta$ lines. The spectra were normalized to the
neighboring continuum. }
\end{figure}

\subsection{H$\alpha$ and H$\beta$ Equivalent width evolution}

Shortly after eruption, novae develop strong and broad ($\geq1000$
km s$^{-1}$ H$\alpha$ emission lines that fade slowly (Shafter \&
Irby 2001). Figure 5 displays the H$\alpha$ and H$\beta$ emission
lines of the M~31 nova, and their evolution during the period of the
observations. Both lines show rather asymmetric saddle shaped
profiles, with the blue component more prominent than the red. As
the nova evolves the line width decreases while their intensity
increases and the double peak profile becomes more distinct. Very
similar behavior of the H$\alpha$ and H$\beta$ emission lines has
been observed during the early decline phase of Nova V382 Vel 1999
(Della Valle et al. 2002).

 Table 4 gives the equivalent width (EW) and full width at half
maximum (FWHM) of the H$\alpha$ and H$\beta$ lines.  The main source
of error in the calculation of the equivalent width is the
determination of the continuum. Thus the procedure was repeated five
times, selecting different continuum points in each case. The
average values of the derived equivalent width and the associated
standard deviations are reported in Table 4. They range from 550\AA\
to 1100 \AA\ for H$\alpha$ and from 160\AA\ to 210 \AA\ for
H$\beta$.

The FWHMs were corrected for instrumental broadening according to
the position of the line in the CCD chip by using a near-by arc
line. The evolution of the FWHM during the period of our
observations is shown in Figure 6. This early decline of the
expansion velocity of the nova follows a power law in time  with an
exponent of $-0.19\pm0.04$ for the H$\alpha$ line, and of
$-0.28\pm0.03$ for the H$\beta$ line, which are comparable to the
values found for the early decline of Nova V382 Vel 1999 (Della
Valle et al. 2002).

\begin{figure}
\resizebox{\hsize}{!}{\rotatebox{0}{\includegraphics{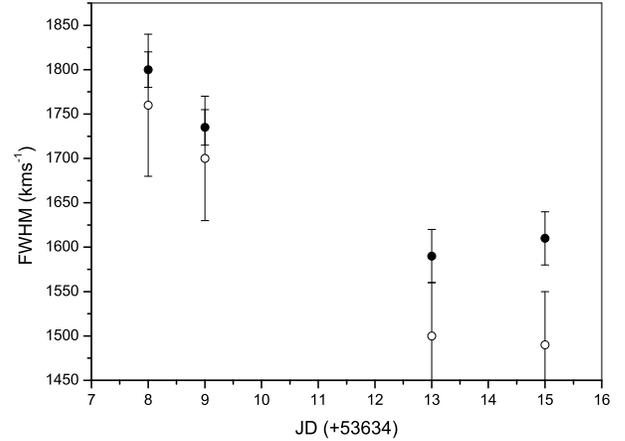}}}
\caption[]{The evolution of the FWHM of the H$\alpha$ (filled
circles) and H$\beta$ lines (open circles). JD reference is 53634}
\end{figure}

\begin{table}
\begin{center}
\caption{Equivalent widths of H$\alpha$ and  H$\beta$ emission
lines } \label{ew}
\begin{tabular}{cccccc}
\hline\hline\noalign{\smallskip}
Date &  MJD  & \multicolumn{2}{c}{EW (\AA)}& \multicolumn{2}{c}{FWHM$^1$(km s$^{-1}$)}  \\
 &   & H$\alpha$  &H$\beta$ & H$\alpha$ &H$\beta$\\
\hline\noalign{\smallskip}
 28sep05&53642&550$\pm$40&160$\pm$5&1800$\pm$20&1760$\pm$80\\
 29sep05&53643&630$\pm$40&160$\pm$5&1735$\pm$20&1700$\pm$70\\
 03oct05&53647&900$\pm$100&185$\pm$8&1590$\pm$30&1500$\pm$60\\
 05oct05&53649&1100$\pm$100&210$\pm$10&1610$\pm$30&1490$\pm$60\\
\hline\hline\noalign{\smallskip}
\end{tabular}
\end{center}
\footnotesize{$^1$  The quoted values have been corrected for
instrumental broadening.}

\end{table}

\section{Summary}
We have obtained photometric and spectroscopic observations of
M31N 2005-09c, a classical nova in the disk of M~31, using the
1.3m telescope of the Skinakas Observatory in Crete (Greece),
starting on September 28, 2005, i.e. about 5 days after outburst,
and ending on the October 5, 2005, i.e. about 12 days after
outburst. We also used supplementary photometric observations
from the La Sagra Observatory in Northern Andaluc\'{\i}a, Spain,
on September 29 and 30, October 3, 6 and 9 and November 1, 2005.
The nova displays a typical early decline spectrum that is
characterized by many weak \ion{Fe}{ii} multiplet emissions. It is
classified as a P$_{fe}$ nova. From the nova light curve, we have
also derived its speed class, $t_2=14\pm2.5 days$. The full width
at half maximum  of the H$\alpha$ and H$\beta$ lines is $\simeq
1700$ kms$^{-1}$ and $\simeq 1600$ kms$^{-1}$, respectively . The
early decline of the expansion velocity of the nova follows a
power law in time with an exponent of $\simeq-0.2$.

\acknowledgements We thank the referee, Dr. A. Evans, whose comments and
suggestions improved the original version of this paper. The authors are
thankful to T. Koutentakis  who helped with the observations at Skinakas
Observatory and Nicol\'{a}s Morales for observations at the (OAM) La Sagra
observatory. Sincere thanks are also due to Dr R. Quimby for providing us
with additional information regarding the nova photometry given in Quimby
et al. (2005).

\end{document}